\documentclass[a4paper, 12pt]{article}
\usepackage{amsfonts}
\usepackage{amsmath}
\usepackage{amssymb}
\usepackage{graphicx}%
\usepackage{color}
\usepackage{float}
\usepackage{geometry}
\usepackage[numbers,sort&compress]{natbib}

\title{Gravitational Wave-Sensitive Photonic-Like Electronic Transport in Graphene for Efficient High-Frequency Gravitational Wave Detection}

\author{\small Shen Shen$^{1,2,3}$, Liangzhong Lin$^{4}$, Linfu Li$^{1}$, Jiang-Tao Liu$^{1*}$,  Xin Wu$^{5,6*}$, Zhenhua Wu$^{2*}$\\
\footnotesize
$^{1}$School of Physics and Mechatronic Engineering, Guizhou Minzu University, Guiyang, 550025, China\\  \footnotesize  $^{2}$School of Physics, Zhejiang University, Hangzhou 310027, China\\  \footnotesize
$^{3}$The Key Laboratory of Optoelectronic Technology \& System, Education Ministry of China, \\ \footnotesize Chongqing University, Chongqing 400044, China \\ \footnotesize
$^{4}$School of Information Engineering, Zhongshan Polytechnic, Zhongshan 528400, China\\ \footnotesize
$^{5}$School of Mathematics, Physics and Statistics, Shanghai University of Engineering Science, \\ \footnotesize Shanghai 201620, China \\ \footnotesize $^{6}$School of Mathematics, Physics and Statistics \& Center of Application and Research of \\ \footnotesize  Computational Physics,
Shanghai University of Engineering Science, Shanghai 201620, China \\
 \footnotesize $^{*}$Email: jtliu@semi.ac.cn;  wuxin\_1134@sina.com;  wuzhenhua@zju.edu.cn
 }

\begin{document}

\maketitle
\begin{abstract}
High-frequency gravitational waves are crucial for understanding the very early universe and distinguishing between various cosmological models, but detecting them remains a significant challenge. We investigated the effects of high-frequency gravitational waves on photonic-like electronic transport in graphene. The results show that, unlike the influence of gravitational waves on the propagation of light, the influence of gravitational waves on photonic-like electronic transport can accumulate not only in real space but also in $k$-space. This makes photonic-like electronic transport under gravitational waves similar to the propagation of light in a medium where the refractive index varies dramatically due to gravitational waves, and with shorter wavelengths. As a result, the relative intensity variation in photonic-like electronic transport under gravitational waves exceeds that of a laser interferometer with the same arm length by six orders of magnitude. At low temperatures, the influence of phonons on photon-like transport in the context of high-frequency gravitational waves can be ignored. These findings indicate a strong interaction between gravitational waves and electron transport, which helps to deepen the understanding of the interaction between gravitational waves and matter, and provides a different method for detecting high-frequency gravitational waves.
\end{abstract}

\section{Introduction}
Gravitational waves are a fundamental prediction of the general relativity theory, and are key to our understanding of gravity, space, and time.
Unlike other forms of radiation, gravitational waves have extreme weak coupling with matter, meaning that they can travel vast distances through the cosmos without being absorbed or scattered,  and can carry rich information about the objects and events that generate them \cite{press1972gravitational,bailes2021gravitational,domenech2021scalar,bian2021gravitational,CTP56Wang_2011}.
Thus, with the advent of advanced high-sensitivity  gravitational wave detectors, researchers may be poised to unlock some of the universe's most profound and enigmatic secrets,  gaining new insights into the nature of black holes, neutron stars, and other exotic phenomena, deepening our understanding of the laws of physics and the evolution of the universe\cite{doi:10.1126/science.abc7397,doi:10.1126/science.abq1187,doi:10.1126/science.abm3231,PhysRevX.13.021020,PhysRevLett.130.181002,PhysRevLett.131.131001,PhysRevX.13.021035}.

While the detection of gravitational waves has been a tremendous success, in early cosmic events such as phase transitions, violent astronomical phenomena, primordial gravitational waves, the merging of primordial black holes, and the intercommutation of cosmic strings\text{-}gravitational waves with frequencies higher than 1000 Hz, and even up to $10^{11}$ Hz, could be generated. As a result, detecting high\text{-}frequency gravitational waves (HFGWs) is extremely important\cite{Aggarwal2021LRR}. However, because the spatial distortions caused by gravitational waves are so minute, detectors need to be several kilometers long to observe them. When the gravitational wave's wavelength is shorter than the detector's length, the detector's sensitivity drops significantly. Current detectors can only capture long\text{-}wavelength (low\text{-}frequency) gravitational waves. Several approaches have been proposed, including ring waveguides\cite{AMCruise_2000}, and detection methods in the EUV and X\text{-}ray bands\cite{Broadfoot1977,10.1093/pasj/59.sp1.S23}. Recently, T. Liu et al. demonstrated that nearby planets, like Earth and Jupiter, could serve as laboratories for detecting high\text{-}frequency gravitational waves \cite{PhysRevLett.132.131402}.

In theory, shortening the wavelength of light could reduce the size of detectors, which would raise the detection frequency. UV-optical, EUV, and X\text{-}ray band detection schemes are based on this idea. However, producing high-quality optical devices in the EUV or X\text{-}ray bands is quite difficult. The wavelength of electrons is much shorter than that of photons. Similarly, as a gravitational wave propagates through a crystal lattice, it causes directional stretching and compression of the lattice, leading to shifts in the electronic energy band. Despite the exciting potential of these changes, few studies have explored this effect in depth, leaving it as an area for future research.

When a gravitational wave arrives, it induces a strain on space-time that elongates it in one direction and compresses it in the direction perpendicular to it\cite{doi:10.1126/science.aat3363}.
This phenomenon has been leveraged by laser interferometry to directly detect gravitational waves, as demonstrated by prior studies~\cite{abbott2009ligo,grote2019novel,campeti2021measuring,mieling2021response,doi:10.1126/sciadv.aau7948,PhysRevX.13.021019,SCP66Li2023}. In addition, atomic interferometers have also been suggested for detecting gravitational waves\cite{doi:10.1126/sciadv.aau7948}.
Similarly, the propagation of a gravitational wave through a crystal lattice would cause a directional stretching and compression of the lattice, leading to changes in  of the electronic energy band.
Despite the intriguing possibilities that could arise from such modifications, few studies have investigated this effect thoroughly, leaving it for future research in depth.

In this study, we investigated the effects of gravitational waves on the electronic band structure, conductivity, and the photonic-like electronic transport behavior of graphene. Graphene, a single layer of carbon atoms arranged in a hexagonal lattice, exhibits unique linear energy dispersion and chiral nature of electrons at the $K$ and $K'$ valleys of the Brillouin zone.
The similarity between the Dirac equation for electrons in graphene and the Helmholtz equation for an electromagnetic wave enables the photonic-like electron propagating behavior in graphene\cite{S315VVC}.
Photonic-like electronic transport refers to the phenomenon the electrons in graphene that resembles the propagation of light.
This unique feature allows for the construction of electronic devices resembling photonic devices, including negative refraction negative refraction\cite{S315VVC,NP11GHL,PRB105TS}, Goos-Hanchen Effect\cite{PRL106WZ}, transformation optics\cite{S332AV}, Quantum Billiards\cite{doi:10.1126/science.1144359}, electronic metamaterial\cite{PRAPP13LX}, Hartman effect\cite{JAP105WZ,NJP14JTL}, and photonic-like highly integrated programmable devices\cite{deng2022graphene}.

Our findings indicate that the spacetime changes induced by gravitational waves primarily alter the reciprocal space of the lattice, modifying the Bloch equation and influencing the band structure and transport behavior of graphene.
The energy band is primarily affected by the neighboring atoms and the strength of gravitational waves is typically small\cite{reich2002tight}, resulting in minimal effects on atomic distance and relative changes in band and conductivity.
To address this limitation, we leveraged the photonic-like electronic transport of graphene under gravitational waves. Since the wavelength of the photonic-like electronic transport is proportional to the Fermi level, the influence of gravitational waves on the Fermi level is the accumulation of  influence of gravitational waves on all electrons in $k$-space below the Fermi level. Furthermore, additional amplified the effects of gravitational waves in real space can be archived by utilizing longer-distance electron transport.
Thus, the accumulation of effects of gravitational waves in both $k$-space and real space significantly amplified the effect of gravitational waves on the photonic-like electron transport, with the intensity variation being six orders of magnitude larger than that of a laser interferometer for the same case. More importantly, higher sensitivity to gravitational waves can be achieved while significantly reducing the size of the detector, and it also allows for the detection of higher frequency gravitational waves. Additionally, since the transport equation for photonic-like electronic transport in graphene aligns with the wave equation for light, existing noise reduction techniques used in optical gravitational wave detectors like LIGO can be directly applied, simplifying detector design.  Therefore, finding suitable methods to detect high-frequency gravitational waves is essential.

\section{Theoretical model and calculation method}

\begin{figure}[H]
\centering
\includegraphics[width=0.9\columnwidth]{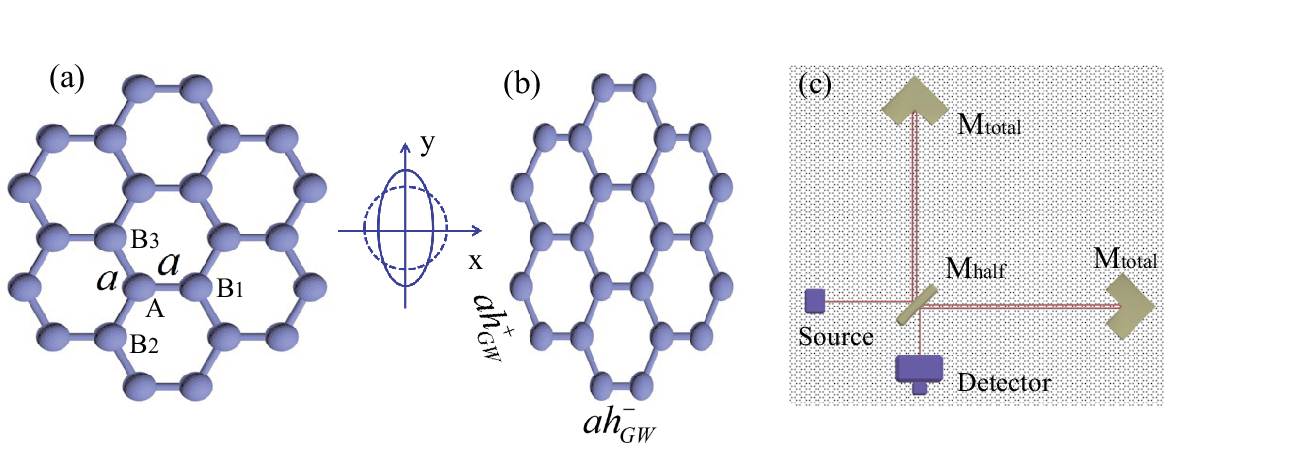}
\caption{(a) The atomic structure of graphene. (b) Schematic diagram of the atomic structure of graphene under the influence of gravitational waves.(b) Schematic diagram of the photonic-like ichelson interferometer.}
\label{fig1}
\end{figure}

Due to the great distance between gravitational wave sources and the Earth, the gravitational waves that propagate to Earth are generally very weak. The spacetime metric caused by gravitational waves can be expressed as adding a small perturbation ${h_{\mu \nu }}$ to a flat spacetime background ${\zeta_{\mu \nu }}$, where ${g_{\mu \nu }}{\rm{ = }}{\zeta _{\mu \nu }}{\rm{ + }}{h_{\mu \nu }}$ is a spacetime metric tensor in general relativity.
In the massless spatiotemporal region, removing the perturbation high-order term can simplify it into a linear wave equation\cite{jiadifferential,Yu-Zhao2012}:
\begin{equation}
\left( {{\nabla ^2} - \frac{1}{{{c_0^2}}}\frac{{{\partial ^2}}}{{\partial {t^2}}}} \right){\tilde h_{\mu \nu }} = 0,\label{eq8}
\end{equation}
where ${\tilde h_{\mu \nu }} = {h_{\mu \nu }} - \frac{1}{2}{\zeta _{\mu \nu }}h$,  $h$ is the contraction of $h_{\mu \nu }$, $c_0$ is the speed of light, then the solution to the eq.~(\ref{eq8}) is:
\begin{equation}
{\tilde h_{\mu \nu }} = h_0{e^{i{k_a}{x^a}}}.
\end{equation}
where $h_0$ is a symmetric $4 \times 4$ constant coefficient matrix, corresponding to the amplitude of each component of the gravitational wave, and ${k_a}$ is a four-dimensional wave vector. If the gravitational waves propagate along the $z$ direction, by introducing the gravitational radiation gauge condition, namely the transverse traceless gauge, the value of $h_0$ can be obtained \cite{jiadifferential}:
\begin{equation}
h_0{\rm{ = }}\left( {\begin{array}{*{20}{c}}
0&0&0&0\\
0&{{h_ + }}&{{h_ \times }}&0\\
0&{{h_ \times }}&{ - {h_ + }}&0\\
0&0&0&0
\end{array}} \right),
\label{eqh0}
\end{equation}
where ${h_{\rm{ + }}}$ is plus polarization and ${h_ \times }$ is cross polarization, they represent two polarization modes of gravitational waves.

The spacetime metric caused by gravitational waves will affect the spatial structure of the crystal lattice, thereby altering the electronic band structure and transport behavior. Before discussing the effects of gravitational waves, we will first present the band structure of graphene without the influence of gravitational waves for comparison. The structure of suspended  graphene consists of two types of non-equivalent carbon atoms, $A$ and $B$, as shown in Fig. 1(a).
The tight binding method can be used to obtain the wave function of an electron as a linear combination of atomic orbital wave functions:
\begin{equation}
\psi \left( {k,r} \right) = c\left| {{\Phi _A}} \right\rangle  + d\left| {{\Phi _B}} \right\rangle,
\end{equation} where $\left| {{\Phi _A}} \right\rangle $ and $\left| {{\Phi _B}} \right\rangle $ are the orbital wave functions of $A$ and $B$ atoms respectively, $c$ and $d$ are undetermined coefficients. Introducing the electron wave function into the energy eigenequation yields:
\begin{equation}
c\hat H\left| {{\Phi _A}} \right\rangle  + d\hat H\left| {{\Phi _B}} \right\rangle  = Ec\left| {{\Phi _A}} \right\rangle  + Ed\left| {{\Phi _B}} \right\rangle,
\end{equation}
where
$\hat H = \frac{{{p^2}}}{{2m}} + \sum\limits_{{R_u}} {V\left( {r - {R_u}} \right)} $ is a Hamiltonian.
For $A$ atom, the coordinates of its nearest neighbors are:
\begin{subequations}
\begin{align}
&{B_1} = a\left( {1,0} \right),\\
&{B_2} = \frac{a}{2}\left( { - 1,\sqrt 3 } \right),\\
&{B_3} = \frac{a}{2}\left( { - 1, - \sqrt 3 } \right),
\end{align}
\end{subequations}
where ${a}$ is the nearest neighbor distance.
Thus, the relationship between graphene energy and wave vector can be obtained by \cite{bena2009remarks}:
\begin{small}
\begin{equation}
E =  \pm {\gamma _{AB}}{\left[ {{\rm{3}} + 4\cos \left( {\frac{3}{2}{k_x}a} \right)\cos \left( {\frac{{\sqrt 3 }}{2}{k_y}a} \right) + 2\cos \left( {\sqrt 3 {k_y}a} \right)} \right]^{1/2}},
\label{band0gw}
\end{equation}
\end{small}where ${{\gamma _{AB}}}$ is the overlapping integral of the nearest neighbors, $q_x$ ($q_y$) are the components of the wave vector in the $x$ ($y$) directions.

Gravitational waves cause the distance between nearest neighbors in graphene to change. Since gravitational waves are transversely unscented, the strength of gravitational waves $h_0$ in Eq. (\ref{eqh0}) is usually expressed by dimensionless amplitude ${{h_{GW}}\left( t \right)}$ \cite{riles2013gravitational}, which is the ratio of the space-time distortion caused by gravitational waves to the flat space-time metric. If the gravitational waves propagate along the $z$ direction, the space is compressed in the $x$ direction and stretched in the $y$ direction. Because of the low frequency of gravitational waves, the energy band in graphene can be approximated as a static process using the adiabatic approximation. At this time, the coordinates of the nearest neighbors are:
\begin{subequations}
\begin{align}
&{B_1} = a\left( {h_{GW}^ - ,0} \right),\\
&{B_2} = \frac{a}{2}\left( { - h_{GW}^ - , - \sqrt 3 h_{GW}^ + } \right),\\
&{B_3} = \frac{a}{2}\left( { - h_{GW}^ - ,\sqrt 3 h_{GW}^ + } \right),
\end{align}
\end{subequations}
  where ${h_{GW}^ + = 1 + {h_{GW}}}$ and ${h_{GW}^ - = 1 - {h_{GW}}}$. Different from lattice deformation caused by stress, the wave function and the distance between atoms changes with the gravitational wave simultaneously, and the change of overlapping integral of nearest neighbors is a higher-order term that can be ignored. And the relationship between energy and wave vector can be written as follows:
\begin{small}
\begin{equation}
E =  \pm {\gamma _{AB}}{\left[ {{\rm{3}} + 4\cos \left( {\frac{{3a}}{2}{q_x}h_{GW}^ - } \right)\cos \left( {\frac{{\sqrt 3 a}}{2}{q_y}h_{GW}^ + } \right) + 2\cos \left( {\sqrt 3 a{q_y}h_{GW}^ + } \right)} \right]^{1/2}}.
\label{band1gw}
\end{equation}
\end{small}
By comparing Eq. (\ref{band0gw}) and Eq. (\ref{band1gw}), it can be observed that gravitational waves alter the distances between carbon atoms in graphene, thereby changing its lattice structure and causing a slight shift in the electron wave vectors. This will affect the electronic transport behavior.

In this paper, to better reveal the effects of gravitational waves on electronic transport and to make a better comparison with optical gravitational wave detectors, we use the photonic-like transport behavior of graphene. We can also construct a photonic-like interferometer similar to the Michelson interferometer to study the impact of gravitational waves on the photonic-like transport behavior in this structure. Specifically, the interferometer consists of a semi-transparent and semi-reflective mirror ${M_{half}}$, a fully reflective mirror ${M_{total}}$, and a detector [Fig.\ref{fig1}(c)].
The electron beam is divided into horizontal and vertical beams after passing through ${M_{half}}$, and then under the action of ${M_{total}}$, it returns to ${M_{half}}$.
Finally, the two beams of electrons interfere and are received by the detector.
${M_{half}}$ and ${M_{total}}$ are realized by making use of Klein tunneling \cite{katsnelson2006chiral}, by setting an appropriate gate.
For instance, when the Fermi energy is 0.1 $eV$, the width of the barrier of ${M_{half}}$ can be set to 43 $nm$, and the height of the barrier can be set to 0.025 $eV$, achieving a transitivity and reflectivity of around 0.5.
${M_{total}}$ is designed as a right-angle mirror with an included angle of ${90^\circ}$, and the incidence angle of electrons is ${45^\circ}$.
At this point, the barrier width can be set to 65 $nm$, and the barrier height can be set to 0.117 $eV$, allowing for a photonic-like interferometer to be created.

Because the motion of electrons during photonic-like transport processes is consistent with the propagation of light waves, the methods for studying the interference and output of electrons in photonic-like transport processes are the same as those in a laser Michelson interferometer \cite{born_wolf_bhatia_clemmow_gabor_stokes_taylor_wayman_wilcock_1999}. In a photonic-like interferometer, destructive interference between two electron beams at the detector can be achieved when the lengths of the horizontal arm ${{l_x}}$ and the vertical arm ${{l_y}}$ satisfy the relationship: ${{l_x} = \frac{{2{l_y}{q_y} - \left( {2n + 1} \right)\pi }}{{2{q_x}}}\left( {n = 0,1,2, \cdots } \right)}$. While constructive interference occurs when they meet the relationship: ${{l_x} = \frac{{2{l_y}{q_y} - 2n\pi }}{{2{q_x}}}\left( {n = 0,1,2, \cdots } \right)}$.
And the phase difference between the two electron beams at the detector caused by gravitational waves is:
\begin{equation}
\Delta \phi  = \Delta \phi_{al}+\Delta \phi_{k}+O(1),
\end{equation} where $ \Delta \phi_{al}=2{{h_{GW}}({l_y}{q_{{y_0}}} + {l_x} {q_{{x_0}}}})$
 is caused by the change of the arm's length by the gravitational wave,
$\Delta \phi_{k}={2{l_y}\Delta {q_y}- 2{l_x}\Delta {q_x}}$ is caused by the change of wave vector under the action of gravitational wave, and $O(1)$ are high-order small quantities that can be ignored.

\section{Numerical  results}
\begin{figure}[H]
\centering
\includegraphics[width=0.95\columnwidth]{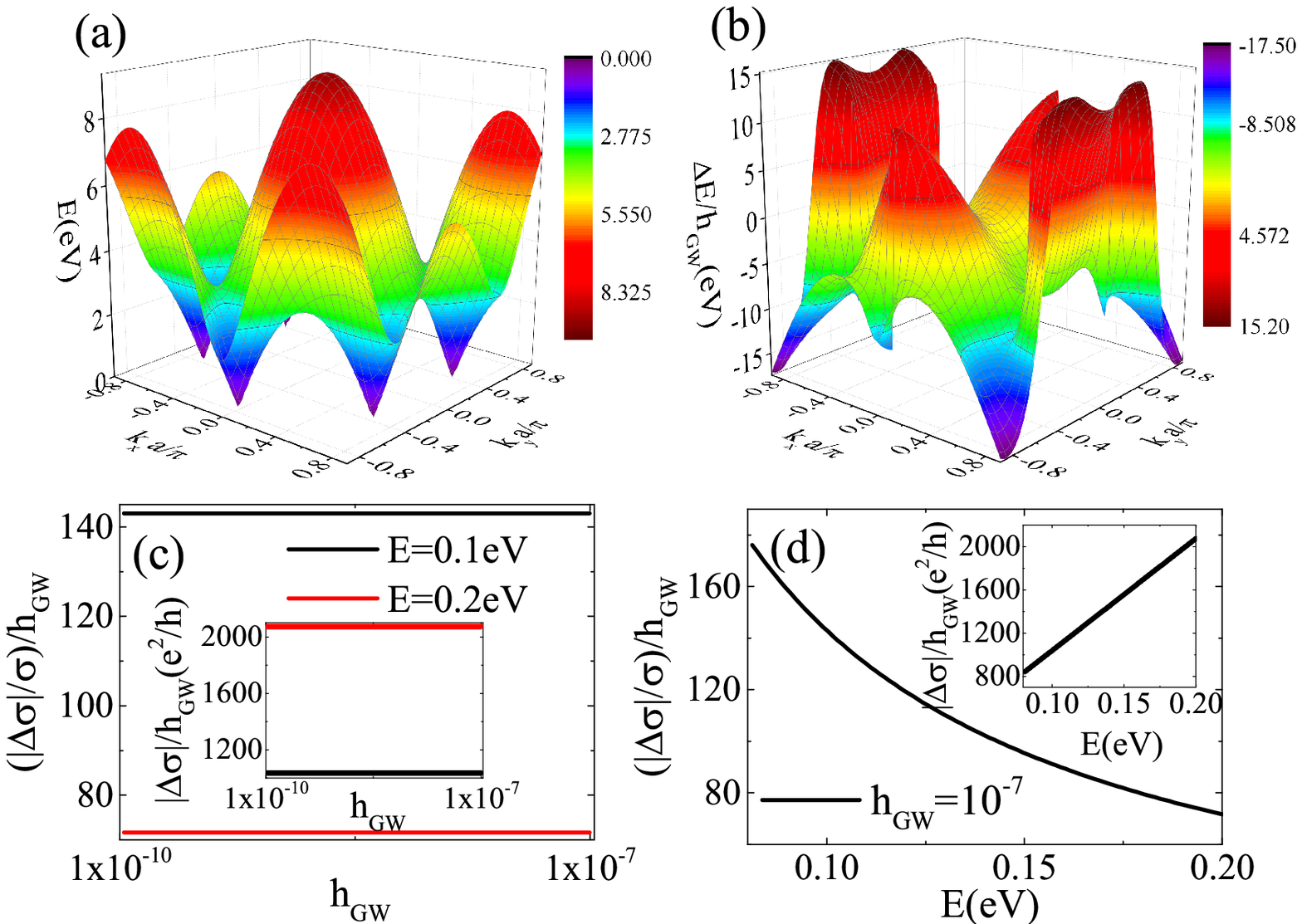}
\caption{(a) Conduction band diagram of graphene under the action of gravitational waves. (b) Ratio of the electron energy change and ${{h_{GW}}}$ under the effect of gravitational waves. The relative change of the ratio of graphene conductivity to ${{h_{GW}}}$ varies with (c) ${{h_{GW}}}$ and (d) Fermi energy. The illustrations show the absolute change in conductivity with (c) ${{h_{GW}}}$ and (d) Fermi energy, respectively.}
\label{fig2}
\end{figure}

The energy band of graphene affected by the gravitational waves is shown in Fig. \ref{fig2}.
Despite the slight shift in the position of the 6 Dirac points caused by the gravitational waves\cite{PhysRevB.99.064302}, the energy variation near the Dirac points remains nearly linear with respect to the wave vector.
For example, when ${{q_x} = 0}$, the ${{q_y}}$ of Dirac points shift to ${ \pm \frac{{4\pi }}{{3\sqrt 3 a {h_{GW}^ +} }}}$.
However, the gravitational waves alter the distance between atoms, causing a slight shift in the energy bands, with a maximum change of 15 ${h_{GW}}$ in energy levels [Fig. \ref{fig2}(b)]. The change in the wave vector of an individual electron is proportional to the dimensionless amplitude of the gravitational wave. In graphene, the energy of electrons is roughly linearly related to their wave vector. Consequently, the relative energy change of electrons induced by gravitational waves ($\Delta E/E/ h_{GW}$) is approximately equal to 1.
Although the relative change in band shift is larger than that in the spatial change ${h_{GW}}$, there is no significant increase in magnitude.
Consequently, detecting the resulting band shift in graphene due to small gravitational wave amplitudes ${{h_{GW}}}$ is challenging.

To gain a better understanding of how gravitational waves affect carriers and to make changes in energy bands more observable, we examined their impact on the electrical conductivity based on studies of the effect of gravitational waves on the energy band.
The conductivity of graphene can be expressed as ${\sigma = ne\mu}$ \cite{chen2008charged,ando2006screening,trushin2007minimum}, where ${\mu}$ is the mobility of carriers \cite{bolotin2008ultrahigh,chen2008intrinsic}, $e$ is quantity of electric charge, and ${n}$ is the carrier concentration.
The conductivity increases with an increase in carrier concentration, which can be calculated by the density of energy states $g(E)=\frac{g_{s}g_{v}S}{\left(2\pi\right)^{2}}\int\frac{dl}{\left\rceil \nabla_{q}E\left(q\right)\right\lceil }$, where ${g_v}$ is the valley degeneracy and ${g_s}$ is the spin degeneracy, $S$ is the area of graphene.

Under the influence of gravitational waves, both the energy band and the density of energy states are altered.
Therefore, the carrier concentration in graphene changes under the influence of gravitational waves, even through the Fermi energy remains unchanged.
When the electron energy is 0.1 $eV$, the relative change in conductivity is approximately 100 times greater than that in the spatial change ${h_{GW}}$.
It is also much higher than the relative change in the energy band.
This is because the change of conductivity is related to not only the change in the energy band but also the corresponding change in the density of energy states.
This represents the impact of gravitational waves on all electrons, which leads to an accumulation of electrons in $k$-space and amplifies the effects of gravitational waves.
As the change in carrier concentration in graphene increases almost linearly with the increase in ${h_{GW}}$, the conductivity of graphene also increases almost linearly with ${h_{GW}}$ [Fig. \ref{fig2}(c)].
The ratio of the relative and absolute changes in graphene conductivity (the inset of Fig. \ref{fig2}(c)) to ${h_{GW}}$ is roughly constant.
Thus, we choose a larger gravitational wave radiation intensity for calculations to reduce the impact of computational errors.

In addition, the number of electrons impacted by gravitational waves also rises as the Fermi level increases.
Consequently, the absolute change in graphene conductivity caused by gravitational waves increases.
However, the relative change in graphene conductivity influenced by gravitational waves decreases as the Fermi level increases (Fig. \ref{fig2}(d)).
This is because the rate of change in graphene conductivity resulting from gravitational waves is smaller than the increase in conductivity.

Although the relative change in conductivity is two orders of magnitude larger than that in the spatial displacement ${h_{GW}}$, detecting small changes in conductivity caused by weak gravitational waves remains challenging.
To enhance the absolute change in resistance or current induced by gravitational waves, the length of the graphene can be increased.
However, this does not lead to an increase in the relative changes in resistance or current.
As the radiation intensity of gravitational waves is small and the atomic distance in graphene is also tiny, the change in atomic distance caused by gravitational waves is minimal.
According to the tight binding approximation, the energy band structure of electrons is only marginally affected by the distance between neighbors, leading to a minor impact on the energy band and corresponding conductivity under the influence of gravitational waves.
To detect gravitational waves, laser interferometers use the interference of ultra-long distance light to amplify their effects.
Therefore, we studied the impact of gravitational waves on the long-range photonic-like electron transport process in graphene.

\begin{figure}[H]
\centering
\includegraphics[width=0.9\columnwidth]{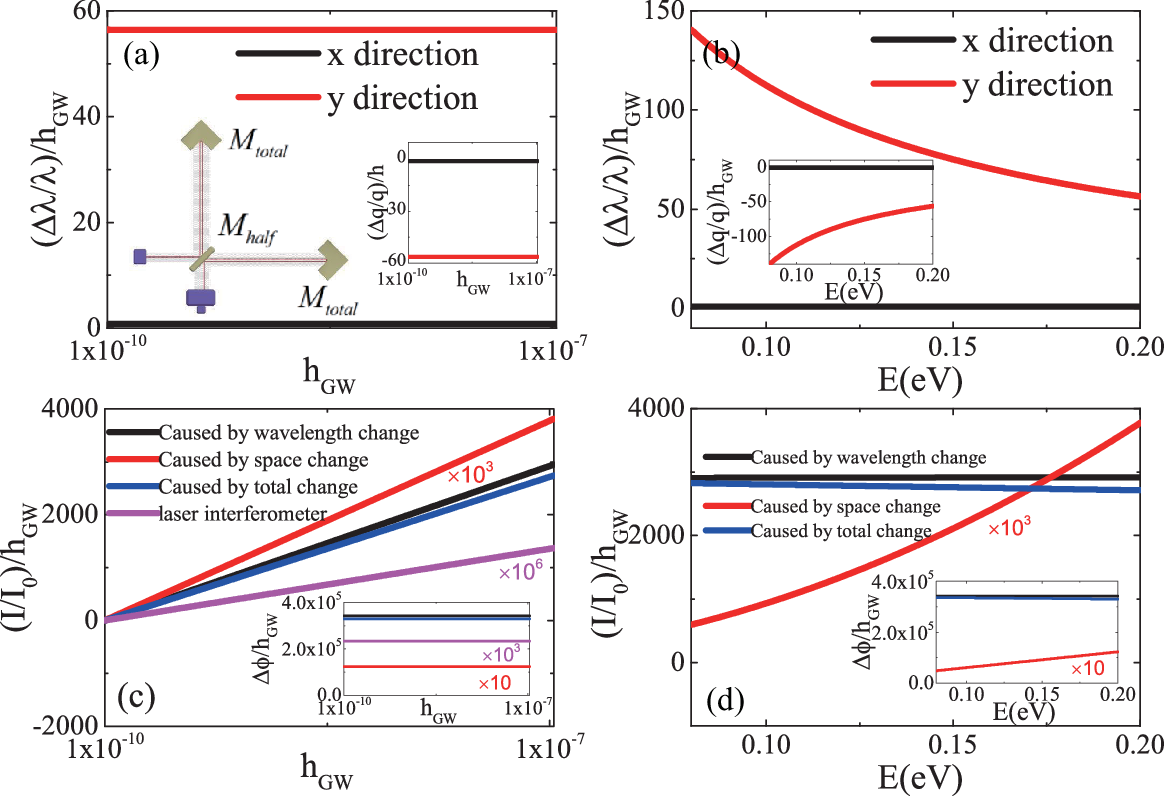}
\caption{The ratio of the change in electronic wavelength to ${{{h_{GW}}}}$ as a function of (a) gravitational wave radiation intensity ${{{h_{GW}}}}$ and (b) Fermi energy level, with the inset showing the ratio of the change in wave vector to ${{{h_{GW}}}}$ as a function of (a) gravitational wave radiation intensity ${{{h_{GW}}}}$ and (b) Fermi energy level. The relative change in electron beam intensity received by the detector and the ratio to ${{{h_{GW}}}}$ as a function of (c) gravitational wave radiation intensity ${{{h_{GW}}}}$  and (d) Fermi energy level. The ratio of the phase difference between the two electron beams at the detector to ${{{h_{GW}}}}$ as a function of (c) gravitational wave radiation intensity and (d) Fermi energy level.}
\label{fig:3}
\end{figure}

To take the advantage of unique photonic-like electronic transport behavior in graphene, we propose a design of a photonic-like interferometer (Fig. 1(c)).
The changes in wave vector or wavelength induced by gravitational waves are depicted in Fig. \ref{fig:3} (a) and (b).
When the polarization direction of the gravitational wave is along the $z$-axis, the $y$-direction lattice of the photonic-like interferometer is stretched while the $x$-direction lattice is compressed.
This alteration in the energy band leads to a decrease in the wave vector in both the horizontal and vertical directions, with a greater change observed in the $y$-direction compared to the $x$-direction.

As the $h_{GW}$ increases, the relative change in wave vector and  wavelength increases, but the ratio of the relative change in wave vector and wavelength to $h_{GW}$ remains constant (Fig.\ref{fig:3}(a)).
The relative change in wavelength is about two orders of magnitude larger than that in spatial change $h_{GW}$.
This is because the change in Fermi energy is related to the shift of the energy band and the corresponding change in the density of energy states, which leads to an accumulation of effects of gravitational waves on electrons in $k$-space.
When the Fermi energy increases, the relative changes in the wave vector and wavelength decreases (Fig. \ref{fig:3}(b)).
In both cases, the variation in wavelength and wave vector in the vertical direction is always greater than that in the horizontal direction. In traditional optics, a change in the index of refraction will also result in a change in wavelength. Thus, the photonic-like interferometer can be regarded as a traditional laser interferometer whose refractive index changes sharply with the gravitational wave.

Fig.\ref{fig:3}(c) shows the phase change and output current change between the two arms of the photonic-like interferometer caused by gravitational waves with 10 $\mu$m arm length, and the expression for relative intensity is: ${I \mathord{\left/
 {\vphantom {I {{I_0}}}} \right.
 \kern-\nulldelimiterspace} {{I_0}}} = {{\left( {1 + \cos \Delta \phi } \right)} \mathord{\left/
 {\vphantom {{\left( {1 + \cos \Delta \phi } \right)} 2}} \right.
 \kern-\nulldelimiterspace} 2}$, where $I_{0}$ denotes the maximum intensity when the two beams of electrons interfere constructively.
The ratio of the relative change of the wave vector to $h_{GW}$ is constant, as is the ratio of the gravitational wave to the change of the arm's length.
Consequently, the ratio of the phase difference to $h_{GW}$ is approximately constant.
As the impact of the gravitational waves on the relative and absolute changes of phase increases with distance, the influence of gravitational waves on phase can be further accumulated in space, greatly improving sensitivity to detect gravitational waves.
The phase difference caused by the change of wave vector is about 4885 times larger than that due to the change of the arm's length, demonstrating that the accumulation of electrons in $k$-space and real space can significantly amplify the effect of gravitational waves.

 For comparison, we have included the phase and output current changes between the two arms of a laser interferometer caused by gravitational waves as shown by the purple line in Fig. \ref{fig:3}(c).
At a typical laser wavelength of 1064.5 $nm$, the relative intensity change due to the influence of gravitational waves is $6$ orders of magnitude smaller than that in the graphene photonic-like interferometer with the same arm length.
Note that, the relative intensity change caused by arm length change in the photonic-like interferometer is about 2782 times larger than that in the laser interferometer, which is due to the shorter electron wavelength in the photonic-like electronic transport.
As wave vector changes accordingly due to the gravitational wave, superior high sensitivity can be achieved in the proposed graphene based photonic-like interferometer. Thus, the size of the gravitational wave detector based on photonic-like transport can be much smaller, and it can detect higher frequency gravitational waves.

Additionally, when the Fermi level of graphene increases, the values of ${q_{x}}$ and ${q_{y}}$ increase, resulting in an increase in the phase change caused by the change of arm's length.
However, the change in wave vector caused by gravitational waves varies slightly with the Fermi energy of graphene (as shown in the inset of Fig. \ref{fig:3}(d)).
Since the polarity of the phase difference caused by the wave vector change and the arm length change is opposite, the total phase difference decreases with the increase of the Fermi energy level.

\begin{figure}[H]
\centering
\includegraphics[width=0.95\columnwidth]{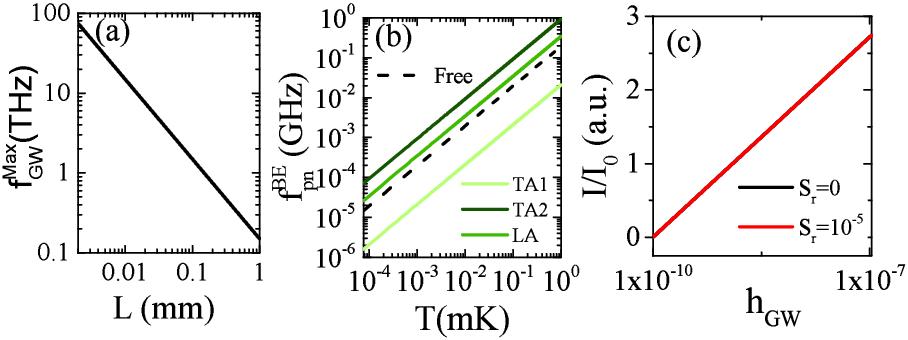}
\caption{(a) The relationship between detector length and the maximum frequency of gravitational waves. (b) The variation of characteristic frequencies of different phonons with temperature under the Bose-Einstein distribution. (c) The ratio of the change in electronic wavelength as a function of gravitational wave radiation intensity under different static strains in graphene.}
\label{fig:4}
\end{figure}

As the Big Bang theory continued to develop and improve, the inflationary model of the universe was established. This model predicts a rapid inflationary period following the Big Bang, which would lead to the production of high-energy relic gravitational waves. The frequency range of these relic gravitational waves extends from approximately $~10^{-18}$ Hz to a high-frequency cutoff of $10^{10}$ Hz\cite{Aggarwal2021LRR,n460Abbott2009,GASPERINI20031}. Additionally, the merging of primordial black holes and the intercommutation of cosmic strings can also result in high-frequency gravitational wave radiation. However, when the half-wavelength of the gravitational wave is smaller than the length of the detector, the detector's sensitivity decreases sharply. Traditional gravitational wave detectors, which are longer in length, can only detect lower frequency, longer wavelength gravitational waves. Due to the photon-like transport properties in graphene being more sensitive to gravitational waves, the length of the detector can be reduced. With a detector arm length of 1 mm, gravitational waves with a maximum frequency of 0.15 THz can be detected [Fig. \ref{fig:4}(a)].

Furthermore, phonons can also cause lattice distortions, thereby affecting the detection results. However, gravitational wave detection can be conducted at extremely low temperatures. For example, in the next generation of detectors, mirror temperatures can be lowered to 77 nk\cite{doi:10.1126/science.abh2634}. Since phonons obey the Bose-Einstein distribution, the characteristic phonon energy is equal to $k_{B}T$, where $k_{B}$ is the Boltzmann constant, $T$ is the temperature. The characteristic phonon energy varies with temperature as shown in Fig. \ref{fig:4}(b). The calculations took into account the TA1 phonons, TA2 phonons, and LA phonons in graphene, as well as free phonons \cite{https://doi.org/10.1002/sia.1948}.  At a temperature of 1 mK, the highest energy TA1 phonon's  characteristic frequency is only 9.2 MHz. If the gravitational wave frequency is much higher than the characteristic phonon frequency, for instance, if the gravitational wave frequency exceeds 92 MHz, the impact of phonons can be entirely ignored. At a temperature of 77 nK, the highest energy TA1 phonon's characteristic frequency is only 71 KHz, allowing the detectable gravitational wave frequency to be further reduced.

Due to stress or gravity, there might be some initial axial stress or strain present in the graphene. Although strain can affect electron transport, this influence is static and does not cause the interference current to vary over time, thus not affecting the detection of gravitational waves. In Fig. \ref{fig:4}(c), we studied the deformation of graphene caused by external forces, with a deformation of approximately $10^{-5}$. The change in interference current under the influence of gravitational waves is consistent with the change in interference current when there is no deformation.

Finally, while this paper is to uncover the physical mechanisms of the impact of gravitational waves on electronic transport and to elucidate why this impact is significantly greater for electronic transport than for photonic transport. It delves into novel physics mechanisms without the primary objective of designing a gravitational wave detector. Although we touch upon the feasibility of experimental implementation, it remains outside the scope of this study. We also note that, the detection of gravitational waves using photonic-like electronic transport in graphene is still a challenging task.

Thanks to its high sensitivity to gravitational waves, photonic-like electronic transport has the potential to significantly reduce the size of gravitational wave detectors by six orders of magnitude, which implies that the size of the graphene photonic-like interferometer is small in size, making it possible to develop desktop or even chip-level gravitational wave detectors.
Like optical gravitational wave detectors, external vibrations can induce deformation in the carbon lattice, leading to noise.
It is important to note that traditional optical gravitational wave detectors, such as the mirrors in LIGO, have heavy masses (up to 40 kg) and optical circuit lengths of several kilometers, making vibration isolation extremely challenging. In contrast, the system size proposed in this paper is on the millimeter scale, making vibration isolation much simpler. Similarly, temperature noise is present in traditional optical gravitational wave detectors, but in the next generation of detectors, mirror temperatures can be lowered to 77 nanokelvin\cite{doi:10.1126/science.abh2634}. Although the impact of phonons can be ignored at higher gravitational wave frequencies, for low-frequency gravitational waves, the influence of phonons still needs to be carefully studied. However, theoretically, it is possible to reduce or eliminate the impact of phonons through coherent detection using two detectors. For example, at extremely low temperatures, it is difficult for characteristic phonons with similar amplitudes and frequencies to simultaneously appear in different detectors, correlation can be used to eliminate characteristic phonon-induced noise. This technique is well-established in current gravitational wave detection\cite{Rowan2016,PhysRevD.82.103007}. Furthermore, traditional optical gravitational wave detectors have mirrors weighing over 10 kg, whereas the proposed system size in this paper is small, and the thickness of the two-dimensional materials is extremely thin, making it easier to achieve very low temperatures compared to traditional optical gravitational wave detectors.

Furthermore, the black holes can be generated in heavy ion collision experiments\cite{hanauske2019magic,bernitt2022fundamental}, which can also radiate gravitational waves in a laboratory setting.
Although the intensity of gravitational waves produced by these collisions is lower than that of celestial bodies, the detection distance of micro gravitational wave detectors is much smaller, making it possible to detect gravitational wave radiation in these experiments.
Additionally, the initial velocity of the colliding particles in heavy ion collisions is higher, which leads to stronger gravitational wave radiation\cite{PhysRevLett.101.161101}.
Therefore, as the energy of heavy ion collisions increases, it may become feasible to use miniature gravitational wave detectors to detect gravitational wave radiation in these experiments in the future.

\section{Conclusions}
In this study, we have investigated the impact of gravitational waves on electron transport in graphene and designed a photonic-like interferometer that utilizes the unique linear energy dispersion and the chiral nature of electrons in graphene. The results  indicate that the spacetime changes induced by gravitational waves primarily alter the reciprocal space of the lattice, modifying the Bloch equation and influencing the band structure and transport behavior of graphene.  Since the wavelength of the photonic-like electronic transport is proportional to the Fermi level, the influence of gravitational waves on the Fermi level is the accumulation of  influence of gravitational waves on all electrons in $k$-space below the Fermi level. Furthermore, additional amplified the effects of gravitational waves in real space can be archived by utilizing longer-distance electron transport.
Thus, by accumulating the effect of gravitational waves in both real space and $k$-space,  the interferometer can be regarded as an traditional laser interferometer whose refractive index changes drastically with gravitational waves, so its sensitivity has been significantly improved.
The results demonstrate that, under similar conditions, the photonic-like interferometer is $6$ orders of magnitude more sensitive than a laser interferometer in detecting the intensity change caused by gravitational waves. This could significantly reduce the size of detectors, making it possible to detect high-frequency gravitational waves.   These findings indicate a strong interaction between gravitational waves and electronic transport, which contributes to a deeper understanding of the interaction mechanisms between gravitational waves and matter and provides a different method for detecting high-frequency gravitational waves.

\section*{Acknowledgments}
This work was supported by National Natural Science Foundation of China (NSFC) (Grants No. 62174040, No. 12174423),  the 13th batch of outstanding young scientific and Technological Talents Project in Guizhou Province [2021]5618, Provincial Science and Technology Projects
of Guizhou ZK[2024]501, and Project supported by the Scientific Research Fund of Guizhou Minzu University (GZMUZK[2023]CXTD07), the College Innovation Project of
Guangdong Province (Grant No. 2022KTSCX341), and the MOST (Grants No. 2021YFA1200502).


\end{document}